# On Achievable Rates for Non-Linear Deterministic Interference Channels

Amin Jafarian and Sriram Vishwanath
Department of Electrical & Computer Engineering
University of Texas, Austin, USA
Email: {jafarian, sriram}@ece.utexas.edu

*Abstract*—This paper extends the literature on interference alignment to more general classes of deterministic channels which incorporate *non-linear* input-output relationships. It is found that the concept of alignment extends naturally to these deterministic interference channels, and in many cases, the achieved degrees of freedom (DoF) can be shown to be optimal.

## I. INTRODUCTION

There is a growing body of work on achievable rate regions for multiuser interference channels (M-IFC). The primary difference in literature between the two-user interference channel and the more-than-two-user is the notion of alignment [5]. In a $K$-user multiuser interference channel, the $K-1$ sources of interference at any one receiver are aligned to minimize impact to the legitimate signal.

This concept of alignment has been applied to a fairly wide class of M-IFCs, including the time-varying Gaussian interference channel [1], the static Gaussian interference channel [2], [4], [9] and finally, the deterministic approximation of a Gaussian interference channel [6]. Although a general theory for alignment for memoryless channels remains elusive, there has been limited success in generalizing alignment and applying the concept to some additional classes of deterministic interference channel models. In [8], the authors consider a class of separable deterministic interference channels and develop random-coding based alignment schemes for them. Alignment schemes for a class of cyclically symmetric deterministic interference channels are analyzed in [7]. In spite of this body of work, we have yet to gain a good understanding of alignment and its value, even for deterministic interference channel models.

In this paper, we take a significant step in determining the applicability and implications of alignment by analyzing classes of non-linear deterministic interference channels. Unlike the implicit characterizations of deterministic channels used in [3], [7], we analyze three classes of deterministic channels with an explicit functional form (see Section V). This enables us to develop constructive algebraic arguments for alignment schemes and gain an intuitive understanding of alignment and its repercussions for general IFCs.

The rest of this paper is organized as follows: the next section presents the channel model. Definitions and a limited

This work was conducted with the suppport of AFOSR under Grant FA95500910063, and the NSF under Grants CCF-0905200 & -CCF-0916713.

background is provided in Section III. A general upper bound on degrees of freedom for the class of interference channels studied in this paper is presented in Section IV. Next, alignment schemes for three classes of non-linear deterministic channels are presented in Section V. Finally, the paper concludes with Section VI.

## II. DETERMINISTIC CHANNEL MODEL

Our focus of study is the symmetric $K$-user discrete deterministic channel, where the channel model is given as:

$$Y_i = h(X_i, X_{[i+1] \bmod K+1}, ..., X_{[i+K-1] \bmod K+1}), \quad (1)$$

where $h : \mathbb{Z}_q^K \to \mathbb{Z}_q$ for a parameter $q$ that is prime. Equation (1) can be understood as a system where each receiver receives the same function of input signals, circularly symmetric with respect to the transmitters. For all the classes of deterministic interference channels we investigate in this paper, $h(.)$ is a polynomial function whose definition does not depend on the particular value of $q$ (i.e., is the same function for different primes $q$).

## III. PRELIMINARIES AND DEFINITIONS

First, some notation: We denote the set of positive integers as $\mathbb{Z}^+$. For two positive integers $m$, $n$, Let $l$ denote the remainder when $m$ when is divided by $n$. Then $l$ is written as

$$l = [m] \bmod n.$$

If $[a] \bmod p = [b] \bmod p$, then we write $a \stackrel{p}{\equiv} b$ or $p|a-b$.

If User $i$ is associated with the codebook $\mathcal{C}_i \subset \mathbb{Z}_q^l$, the transmission proceeds for the duration of $l$ time slots. The achievable rate at this user is defined as

$$R_i = \frac{1}{l} \log(|\mathcal{C}_i|).$$

Given this, we present some further definitions used in this paper:

**Definition 1** (Invertible Set). *We call a set $\mathcal{C}_f \subset \mathbb{Z}_q$, an invertible set with respect to the polynomial $f(.)$, if:*

$$|f(\mathcal{C}_f)| = |\mathcal{C}_f|.$$

Thus, given a set $\mathcal{C}_f$ is an invertible set with respect to a polynomial $f(.)$, we can find an inverse mapping from $f(\mathcal{C}_f)$ to $\mathcal{C}_f$.

**Lemma 1.** *Let $f(X)$ be a polynomial with degree d. There exists a non-trivial invertible set $\mathcal{C} \subseteq \mathbb{Z}_q$, such that:*

$$\lim_{q \to \infty} \frac{\log(|\mathcal{C}|)}{\log(q)} = 1.$$

*Proof:* Let $GC_i = \{j : f(j) = i\}$, and $Im(f) = \{i : |GC_i| > 0\}$. Note that $|GC_i| \leq d$ for all $i$. Therefore, $|Im(f)| \geq \frac{q}{d}$. We construct an invertible set $\mathcal{C}$ to satisfy the following two conditions:

1) $j \in \mathcal{C}$, if there exists $i \in Im(f)$, and $j \in GC_i$
2) If $j, k \in \mathcal{C}$, there is no index $i \in Im(f)$ such that $j, k \in GC_i$.

In other words, $\mathcal{C}$ corresponds to the set of all the minimum elements $\{j : \exists i \in Im(f), j \in GC_i\}$. It can be checked that $\mathcal{C}$ is invertible and

$$|\mathcal{C}| = |Im(f)| > \left\lfloor \frac{q}{d} \right\rfloor,$$

.  ∎

**Definition 2** (Degrees of Freedom). *We define total degrees of freedom the channel given in (1) as:*

$$\text{DoF} = \limsup_{q \to \infty} \frac{\max\limits_{R_i \text{ is achievable}} \sum_{i=1}^{K} R_i}{\log(q)}. \quad (2)$$

This is similar to the definition of DoF in literature [1]. Note that the supremum is needed in the definition of total degrees of freedom (2) as the achieved rate can fluctuate considerably with the size of the alphabet $q$.

In this work, we use Dirichlet's theorem on prime numbers. For the sake of completeness, we state it below:

**Lemma 2** (Dirichlet's Theorem,[10]). *For any two positive coprime[1] integers $a$ and $b$, there are infinitely many primes, $q$, of the following form:*

$$q = an + b, \quad n \in \mathbb{Z}^+.$$

### IV. UPPER-BOUND

Again, for the sake of completeness, the remark below presents an obvious upper-bound on total degrees of freedom of the channel given by Equation (1).

**Remark 1.** *Total degrees of freedom is upper bounded by $K$, i.e.,*

$$\text{DoF} \leq K.$$

Next, we improve on this by presenting a better upper bound for a class of deterministic channels given by Equation (1).

**Theorem 1.** *Let $Y_1 = h(X_1, X_2, ..., X_K)$ and other $Y_i$'s be defined symmetrically as in Equation (1). The total degrees of freedom of this channel is upper bounded by $\frac{K}{2}$ if there exists an index $j \in \{2, ..., K\}$, satisfying one of the following:*

- *Condition (1): The degree of $X_j$ in $h(X_1, X_2, ..., X_K)$ is one.*

[1] The greatest common divisor of $a$ and $b$ is 1

- *Condition (2): One can reconstruct $h(X_j, X_{j+1}, ..., X_{[j+K-1] \mod K+1})$, from $h(X_1, X_2, ..., X_K)$, knowing all other $X_i$ $i \neq j$.*

*Proof:* One can only increase the achieved rate by providing additional information (a genie) to each receiver. Thus, we provide $X_j$, $j \neq i$ & $j \neq [i+j-1] \mod K+1$ to Receiver $i$. Realizing that Receiver $i$ must be able to decode $X_i$, the symmetry in the system along with either Condition (1) or Condition (2) implies that Receiver $i$ must also be able to determine $X_{[i+j-1] \mod K+1}$. Under Condition (1), $h(.)$ becomes a linear function of $X_{[i+j-1] \mod K+1}$, and as non-trivial linear functions are invertible, $X_{[i+j-1] \mod K+1}$ can be determined by Receiver $i$. When Condition (2) applies, Receiver $i$ can re-construct Receiver $[i+j-1] \mod K+1$'s signal and thus decode the message $X_{[i+j-1] \mod K+1}$. Thus, we can upper-bound $R_i + R_{[i+j-1] \mod K+1}$ as:

$$R_i + R_{[i+j-1] \mod K+1} \leq \log(q).$$

Writing the similar equation for all $i \in \{1, 2, ..., K\}$ and summing them up, we get:

$$\sum_{i=1}^{K} R_i \leq \frac{K}{2} \log(q),$$

or in the other words:

$$\text{DoF} \leq \frac{K}{2}.$$

This completes the proof. ∎

### V. ACHIEVABILITY: MAIN RESULTS

Unfortunately, determining an achievable rate for an arbitrary channel defined by (1) is a fairly difficult task. Thus, in this section, we identify particular non-linear deterministic interference networks for which a characterization is tractable. Thus, this section has three subsections, one per channel model. For each channel model, we first develop a scheme that achieves a DoF equal to $\frac{K}{2}$, where $K$ is the number of transmitter/receiver pairs present in the network. Next, we detail conditions that need to be satisfied so that the upper bound equals $\frac{K}{2}$. (Note that, although many do, not all channel models in the subsequent subsections satisfy the conditions imposed by Theorem 1.)

#### A. Channel Model I

Consider channels which can be expressed as

$$h(X_1, X_2, ..., X_K) = aX_1^d + h'(X_1, X_2, ..., X_K),$$

for some $d \geq 1$. Moreover, we require that:

The smallest degree term in $h(.)$ is $aX_1^d$ \quad (3)

For this channel, we show that we can achieve a DoF of $\frac{K}{2}$.

**Theorem 2.** *The DoF of channels belonging to Channel Model I (that satisfy (3)) is greater than or equal to $\frac{K}{2}$.*

Next, we present the proof of Theorem 2.

*Proof of Theorem 2:*

Write $d$ as
$$d = 2^t d_0,$$

where $t \geq 0$ is an integer and $d_0$ is the odd divisor of $d$. Let $c$ be the positive integer, which satisfies:
$$c \stackrel{4}{\equiv} 3, \text{ and } c \stackrel{d_0}{\equiv} 1.$$

We know from the Chinese remainder theorem [10] that there always exists such an integer $c$, since $d_0$ and 4 are coprime. Let $p$ be a prime number of the following form:
$$p = (4d_0)n + c \quad n \in \mathbb{Z}^+. \qquad (4)$$

Note that, from Lemma 2, there are infinitely many of such primes $p$. Consider the set of such primes that satisfy $p > a$. Let $q$ be a prime that satisfies the following inequality:
$$p^2 < q < 2p^2.$$

Existence of such prime $p$ follows from Bertrand's theorem [10]. Equivalently, one can rewrite this inequality as:
$$\sqrt{\frac{q}{2}} < p < \sqrt{q}. \qquad (5)$$

We choose to design our code over the field $\mathbb{Z}_q$.

Let $e$ be a primitive root of the prime $p$. We define our codebook as the following:
$$\mathcal{C} = p \times \left[\left\{e^1, e^2, ..., e^{\frac{p-1}{2}}\right\}\right] \mod p. \qquad (6)$$

From Equation (5), we have $\mathcal{C} \subset \mathbb{Z}_q$ and
$$|\mathcal{C}| > \frac{\sqrt{q}}{2\sqrt{2}} - \frac{1}{2}. \qquad (7)$$

Now we present encoding/decoding scheme.

**Encoding:**

This is straightforward - each User $i$ transmits $X_i \in \mathcal{C}$.

**Decoding:**

Receiver $i$ observes:
$$Y_i = aX_i^d + h'(X_1, X_2, ..., X_K).$$

Let $Y_i'$ be defined as the following:
$$Y_i' = [Y_i] \mod p^{d+1}$$

Now, we have:
$$\left[aX_i^d - Y_i'\right] \mod p^{d+1} = 0. \qquad (8)$$

From the codebook construction (6), we know each codeword is a multiple of $p$. Let $X_i = pe^u$, where
$$u \in \left\{1, 2, ..., \frac{p-1}{2}\right\}.$$

We can rewrite Equation (8) as:
$$ae^{ud} \stackrel{p}{\equiv} \frac{Y_i'}{p^d}.$$

since $p > a$, there exists a multiplicative inverse for $a$ that we denote as $a^{-1}$.

We have:
$$e^{ud} \stackrel{p}{\equiv} a^{-1}\frac{Y_i'}{p^d}. \qquad (9)$$

Next, we show that Equation (9) has a unique solution in the set $\{1, 2, ..., \frac{p-1}{2}\}$. We prove this by contradiction. Let $u, v \in \{1, 2, ..., \frac{p-1}{2}\}$ be two answers for Equation (9). From Equation (9), we have:
$$e^{ud} \stackrel{p}{\equiv} e^{vd}. \qquad (10)$$

From the properties of primitive root of a prime, Equation (10) holds if and only if:
$$p - 1 | ud - vd, \qquad (11)$$

or $p - 1 | 2^t d_0 (u - v)$. Utilizing the form of the prime $p$ given by Equation (4), we know $p-1$ and $d_0$ are coprime and the greatest common divisor of $p$ and $2^t$ is at most 2. Therefore we can rewrite Equation (11) to get:
$$\frac{p-1}{2} | u - v,$$

which contradicts the statement that $u$ and $v$ are elements of the set $\{1, 2, ..., \frac{p-1}{2}\}$. Therefore, there is a unique solution for Equation (9), and Receiver $i$ can decode $X_i$ correctly.

∎

Unfortunately, a general converse statement for this class of channels is not possible. The upper bound must be derived in a case by case basis. The following example illustrates the case in which both the achievable scheme and the upper bound give $3/2$ degrees of freedom.

Example 1: Consider the following channel model:
$$h(X_1, X_2, X_3) = f(X_1) + f(X_2)g(X_3) + T(X_3), \qquad (12)$$

where the smallest degree monomial of $T(.)$ has a strictly greater degree than the smallest degree monomial of $f(.)$, and also $g(.)$ has a degree greater than zero. One can check that the model given in Equation (12), satisfies conditions of Theorem 1, and therefore it has at most $\frac{3}{2}$ total degrees of freedom. It is easy to see that it satisfies conditions of Theorem 2. So, total degrees of freedom of channel given by Equation (12), is equal to $\frac{3}{2}$.

Next, we present two other channel classes where $\frac{K}{2}$ total degrees of freedom is achievable.

### B. Channel Model II

Consider the following channel model:
$$Y_1 = f(X_1) + g(X_2, ..., X_K),$$

where the polynomial $g(.)$ satisfies the following:
$$\exists t \in \mathbb{Z}_p, s.t. \ \forall i \in \{2, 3, ..., K\}, \text{ and } \forall x_i \in \mathbb{Z}_p,$$
$$g(X_2, ..., X_{i-1}, t, X_{i+1}, ..., X_K) = 0. \qquad (13)$$

For example, when $K = 2$, $g(x, y) = x.y.m(x, y)$ for all the possible polynomial $m(x, y)$ satisfies the conditions imposed by (13). Specifically, the choice $t = 0$ satisfies (13).

The following lemma characterizes this class of channels more concretely.

**Lemma 3.** *A function $g(.)$ satisfies (13) if and only if*

$$g(X_2, X_3, ..., X_K) = \prod_{i=2}^{K}(X_i - t).m(X_2, X_3, ..., X_K),$$

*for some polynomial $m(.)$.*

*Proof:* Proof is omitted due to space constraints. Intuitively, the proof is a generalization of the $K = 2$ example presented above. ∎

**Theorem 3.** *The degrees of freedom of a channel belonging to the family given by Channel Model II is at least $\frac{K}{2}$.*

*Proof:* In order to prove the theorem, we consider pairs of users, and show that a rate of $\frac{1}{2}$ is achievable for each pair. We obtain a transmit strategy where each transmitter can communicate its message ($\log q$ bits) $X_i \in \mathbb{Z}_q$ in two time slots. Let $\mathcal{C}$ be an invertible set with respect to $f(.)$ as given by Lemma 1. Our coding scheme for this system is given by:

**Encoding:**

*First time slot:* User $i$ transmits its message $X_i \in \mathcal{C}$.
*Second time slot:* User 1 transmits $t$, and for $i \neq 1$, User $i$ repeats $X_i$.

**Decoding:**

*Receiver 1:* The output signal is $Y_{1,1} = f(X_1) + g(X_2, X_3, ..., X_K)$ in the first time-slot and $Y_{1,2} = f(t) + g(X_2, X_3, ..., X_K)$ in the second time-slot. Since $t$ is known, Receiver 1 can compute $f(X_1)$ as

$$X_1 = Y_{1,1} - (Y_{1,2} - f(t)).$$

Since $X_1 \in \mathcal{C}$ and $\mathcal{C}$ is an invertible set with respect to $f(.)$, Receiver 1 can uniquely determine $X_1$ from $f(X_1)$.
*Receiver $i \neq 1$:* The output signal for the second time slot is:

$$Y_{i,2} = f(X_i) + g(X_{[i+1] \bmod K+1}, ..., t, ...,$$
$$X_{[i+K-1] \bmod K+1}).$$

From Equation (13), we have that $Y_{2,2} = f(X_2)$. Since $X_i \in \mathcal{C}$, Receiver $i$ can determine $X_i$ from $f(X_i)$. ∎

Next, we present an example for this case where the achievable and outer bounds on the total degrees of freedom equal $3/2$.

Example 2: Consider the following channel:

$$h(X_1, X_2, X_3) = f(X_1) + X_2 g(X_3), \quad (14)$$

where $f(.)$ and $g(.)$ are given (freely chosen) polynomials. For this case, one can show, from Theorem 1, that the total degrees of freedom is upper bounded by $\frac{3}{2}$. Also, from Theorem 3, one can show that a total degrees of freedom of $\frac{3}{2}$ is achievable. In summary, for the channel model given by Equation (14), the total degrees of freedom equals $\frac{3}{2}$.

*C. Channel Model III*

Consider the following channel:

$$Y_i = X_i \left( \sum_{j \neq i} X_j \right). \quad (15)$$

The following theorem presents an achievable total degrees of freedom for this category of channels.

**Theorem 4.** *The total degrees of freedom of the channel given by Equation (15) is $\frac{K}{2}$.*

The proof of this theorem is considerably more involved than the previous two. Therefore, in this paper, we limit the proof of Theorem 4 for $K = 3$, but the same proof technique extends to all values of $K$. In this case, $Y_i = X_i(X_j + X_k)$ where $\{i, j, k\} = \{1, 2, 3\}$.

*Proof of Theorem 4:* First, note that $\frac{3}{2}$ is trivially an upper bound for the total degrees of freedom. This follows immediately from Theorem 1.

In order to prove the achievability of $\frac{3}{2}$ total degrees of freedom, we show that each user can achieve the symmetric rate point of $\frac{1}{2}$. Let $q$ be a prime greater than 5 that satisfies:

$$q \stackrel{5}{\equiv} 3 \quad (16)$$

Let the codebook $\mathcal{C}$ be defined as

$$\mathcal{C} = \left\{ 20n + 4 \middle| n \in \mathbb{Z}, 0 \leq n \leq \left\lfloor \frac{q-12}{60} \right\rfloor \right\}. \quad (17)$$

One can verify that

$$|\mathcal{C}| > \frac{q-12}{60}. \quad (18)$$

Now we present the encoding/decoding scheme:

**Encoding:**

Each Transmitter $i$ sends $X_i \in \mathcal{C}$ in the first channel use and $X_i + 1$ in the second channel use.

**Decoding:**

Receiver $i$ observes the following two equations:

$$Y_{i,1} = X_i(X_j + X_k), \quad (19)$$

and

$$Y_{i,2} = (X_i + 1)(X_j + 1 + X_k + 1). \quad (20)$$

Let $\alpha = X_j + X_k$. In order to find the value of $t = 2X_i + \alpha$ we can write:

$$Y_{i,2} = (X_i + 1)(2 + \alpha)$$
$$= 2X_i + 2 + X_i\alpha + \alpha$$
$$= 2X_i + \alpha + (2 + Y_{i,1}).$$

Therefore, we can compute $t$ as:

$$t = 2X_i + \alpha = [Y_{i,2} - Y_{i,1} - 2] \bmod q. \quad (21)$$

From the choice of codebook $\mathcal{C}$, we have the following two conditions:
1) $t < q$,
2) $t = 20m + 16 = 4(5m + 1)$.

Combining Equations (19) and (21), $X_i$ should satisfy the following:
$$2X_i^2 + Y_{i,1} \stackrel{q}{\equiv} tX_i. \quad (22)$$

Note that in Equation (22), $Y_{i,1}$ and $t$ both depend on the other transmitters' codewords. Instead, if they were constant, it is fairly straightforward to develop a codebook using which $X_i$ can be uniquely decoded. Given that $Y_{i,1}$ and $t$ are dependent, a more sophisticated design is needed. It is with this in mind that the codebook $\mathcal{C}$ in Equation (17) is defined.

Utilizing condition (2) as presented above, we can simplify Equation (22) to get:
$$2(X_i - (5m+1))^2 \stackrel{q}{\equiv} 2(5m+1)^2 - Y_{i,1},$$

or equivalently,
$$(X_i - (5m+1))^2 \stackrel{q}{\equiv} (5m+1)^2 - \frac{q+1}{2}Y_{i,1}. \quad (23)$$

Let $l$ be the square root of the right hand side of Equation (23), mod $q$, i.e., $l^2 = (5m+1)^2 - \frac{q+1}{2}Y_{i,1}$ and $l < q$. From Equation (23) follows that $X_i$ is one of the following:
$$X_i \stackrel{q}{\equiv} 5m + 1 \pm l. \quad (24)$$

Note that $X_i$, $5m+1$ and $l$ are all positive integers less than $q$. Solving Equation (23), there are two solutions for $X_i$ as follows:
$$X_i = \begin{cases} 5m+1+l & \text{if } 5m+1+l < q \\ 5m+1+l-q & \text{if } 5m+1+l > q \end{cases}, \quad (25)$$

and
$$X_i = \begin{cases} 5m+1-l & \text{if } 5m+1-l > 0 \\ 5m+1-l+q & \text{if } 5m+1-l < 0 \end{cases}. \quad (26)$$

Note that Receiver $i$ can determine $X_i$ only if there is a unique $X_i \in \mathcal{C}$. Next, we show that between the four potential solutions for $X_i$, only one can belong to $\mathcal{C}$. Therefore, this unique solution is the appropriate choice for $X_i$. The construction of the codebook $\mathcal{C}$ in Equation (17) is tailored to make this happen. Note that if $X \in \mathcal{C}$, then $X \stackrel{5}{\equiv} 4$. This, along with the fact that $q$ is a prime number greater than 5 satisfying Equation (16), we define:
$$rem \stackrel{5}{\equiv} \{l+1, l+3, 1-l, 4-l\},$$

as the remainder set of $X_i$ mode 5.

Thus, $X_i$ can be determined uniquely if one of the elements of $rem$ equals 4, and the remainder do not equal 4. Checking this, if
$$l + 3 \stackrel{5}{\equiv} 4,$$

then $1 - l \stackrel{5}{\equiv} 0$, and $2 - l \stackrel{5}{\equiv} 1$. Also if
$$1 + l \stackrel{5}{\equiv} 4,$$

then $1 - l \stackrel{5}{\equiv} 3$, and $4 - l \stackrel{5}{\equiv} 2$. This implies that $X_i$ can be determined uniquely from Equation (24).

Next, we characterize the degrees of freedom of this coding scheme:
$$\text{DoF}_{coding} = \lim_{q \to \infty} \frac{\frac{K}{2} \log |\mathcal{C}|}{\log(q)} \quad (27)$$
$$\geq \lim_{q \to \infty} \frac{\frac{K}{2} \log(\frac{q-12}{60})}{\log(q)} = \frac{K}{2} \quad (28)$$

where Equation (27) follows from the fact that each user achieves a rate $R = \frac{1}{2} \log(\mathcal{C})$ and Equation (28) follows from (17). This concludes the final case of non-linear deterministic channels studied in this paper. ∎

## VI. Conclusion

In this paper, we present achievable schemes for three classes of non-linear deterministic channels. We prove, for each class, that we can achieve $\frac{K}{2}$ total degrees of freedom. We also show that, for a limited set of these channels, $\frac{K}{2}$ is in fact the maximum degrees of freedom.

This work provides us with a framework under which we can better understand alignment for general (non-linear) discrete memoryless channels. Note that the the tools used in this paper are number-theoretic and combinatorial in nature. It is our belief that such tools need to be carefully understood and exploited to develop a more general theory of alignment for arbitrary input-output systems.

## References

[1] V. R. Cadambe and S. A. Jafar, "Interference Alignment and the Degrees of Freedom for the K User Interference Channel", *IEEE Transactions on Information Theory*, Aug 2008, Vol. 54, Issue 8, Pages: 3425-3441.
[2] A. S. Motahari, S. O. Gharan and A. K. Khandani, "Real Interference Alignment with Real Numbers", *submitted to IEEE Transactions on Information Theory*, arXiv:0908.1208.
[3] A. El Gamal and M. Costa, "The Capacity Region of a Class of Deterministic Interference Channels", *IEEE Transactions on Information Theory*, Vol. IT-28, No. 2, pp. 343-346, March 1982.
[4] R. Etkin and E. Ordentlich, "On the Degrees-of-Freedom of the K-User Gaussian Interference Channel", arXiv:0901.1695.
[5] R. Etkin, D. Tse, and H. Wang, "Gaussian interference channel Capacity to within one bit", IEEE Trans. Info. Theory, Dec. 2008.
[6] V. Cadambe, S. A. Jafar, S. Shamai (Shitz), "Interference alignment on the deterministic channel and application to fully connected AWGN interference networks", arXiv:0711.2547, Nov. 2007
[7] B. Bandemer, G. Vazquez-Vilar and A. El Gamal,"On the Sum Capacity of A Class of Cyclically Symmetric Deterministic Interference Channels", *Proc. IEEE Intl. Symp. Inform. Theory*, 2009.
[8] V. R. Cadambe and S. A. Jafar, "Interference Alignment Via Random Coding and the Capacity of a Class of Deterministic Interference Channels", *Proceedings of the 47th Annual Allerton Conference*, 2009.
[9] A. Jafarian, J. Jose and S. Vishwanath, "Algebraic Lattice Alignment for K-User Interference Channels", *Proc. of the 47th Allerton Conference*, 2009.
[10] G. H. Hardy and E. M. Wright, *An Introduction to the Theory of Numbers (5th ed.)*, Oxford University Press, 1980.